\documentclass{moriond}

\def\be{\begin{equation}}
\def\ee{\end{equation}}
\def\bea{\begin{eqnarray}}
\def\eea{\end{eqnarray}}

\def\eg{\emph{e.g.}}     
\def\ie{\emph{i.e.}}     

\usepackage{soul}

\newcommand{\Photo}{}

\begin{document}
	
	\vspace*{4cm}
	\title{Directional detection of Dark Matter with the MIcro-tpc MAtrix of Chambers}
	
	\author{C. COUTURIER\footnote{ccouturi@lpsc.in2p3.fr}, O. GUILLAUDIN, F. NARAGHI, Q. RIFFARD\footnote{\scalebox{.918}[1.0]{Now at APC, Univ. Paris Diderot, CNRS/IN2P3, CEA/Irfu, Obs. de Paris, Sorbonne Paris Cit\'e, 75205 Paris, France}}, D. SANTOS and N. SAUZET}
	\address{LPSC, Universit\'e Grenoble-Alpes, CNRS/IN2P3, 38026 Grenoble, France}
	
	\author{P. COLAS, E. FERRER RIBAS and I. GIOMATARIS}
	\address{IRFU, CEA Saclay, 91191 Gif-sur-Yvette, France}
	
	\author{J. BUSTO, D. FOUCHEZ and C. TAO\footnote{Also at Tsinghua Center for Astrophysics, Tsinghua University, Beijing 100084, China}}
	\address{Aix Marseille Universit\'{e}, CNRS/IN2P3, CPPM UMR 7346, 13288 Marseille, France}
	
	\author{N. ZHOU}
	\address{Department of Physics, CHEP, Tsinghua University, Beijing, China and CICQM, Beijing, China}
	
	\maketitle\abstracts{
		Particles weakly interacting  with ordinary matter, with an associated mass of the order of an atomic nucleus (WIMPs), are plausible candidates for Dark Matter.
		The direct detection of an elastic collision of a target nuclei induced by one of these WIMPs has to be discriminated from the signal produced by the neutrons, which leaves the same signal in a detector. 
		The MIMAC (MIcro-tpc MAtrix of Chambers) collaboration has developed an original prototype detector which combines a large pixelated Micromegas coupled with a fast, self-triggering, electronics. 
		Aspects of the two-chamber module in operation in the Modane Underground Laboratory are presented: calibration, characterization of the $^{222}$Rn progeny.  A new test bench combining a MIMAC chamber with the COMIMAC portable quenching line has been set up to characterize the 3D tracks of low energy ions in the MIMAC gas mixture: the preliminary results thereof are presented. Future steps are briefly discussed.
	}

	\vspace{-6mm}
	\section{Directional detection of Dark Matter}
	
	\vspace{-3mm}
	The motion of stars and clouds of gas in observed galaxies suggests the presence of a spheroidal halo of Dark Matter (DM) that encompasses the galactic disc of visible matter. 
	The density of the local dark matter has been determined, from nearby stars or rotation curves to be between 0.3 and 0.4 GeV/cm$^3$~\citelow{Catena}. 
	In the framework of the WIMP (weakly interacting massive particle) hypothesis, \ie~if we assume Dark Matter interacts weakly with ordinary matter, we can try to look for recoils due to the elastic scattering of these WIMPs: this is called \textit{direct detection}~\cite{Goodman1985}. 
	The kinetic energy of the recoiling nucleus $E_R$ depends on the mass of the WIMP $M_\chi$, its velocity $v$ with respect to the detector, the mass of the target nucleus $M_R$, and the angle of scattering $\theta$ :
	$$E_R = \frac{2\mu^2 v^2 \cos^2 \theta}{M_R}$$
	with $\mu = M_\chi M_R / (M_\chi + M_R)$ the reduced mass of the system \{WIMP - nucleus\}.
	For instance, for a fluorine nucleus $^{19}$F, a WIMP mass of 100 GeV/c$^2$ and a relative WIMP velocity of  300\,km/s, the maximum ($\theta = 0^\circ$) expected kinetic energy of the recoiling $^{19}$F is 26 keV.
	
	The kinetic energy of the recoil is deposited in the detector medium in three channels: ionization, scintillation, heat. 
	Detectors often use two channels  to distinguish  electron recoils from nuclear recoils. 
	Once the electron/nuclear recoil discrimination is efficient enough, the next challenge is to discriminate the direct detection of an elastic collision of a target nuclei on a WIMP from the signal produced by the neutrons; both indeed leave the same energy signal in a detector. 
	Standard, non directional, direct detectors cannot directly distinguish recoils originating from WIMP scattering and those from neutron scattering. They usually estimate the \textit{probability} of the neutron-induced recoils, thanks to Monte Carlo simulations ; however there is a lack of available data for low-energy neutrons.
	As the sensitivity of the direct detectors improves, elastic scattering from neutrinos will represent an even tougher background noise to discern from dark matter recoils~\cite{Billard}.
	The WIMP velocity with respect to the Earth comes primarily from the rotation of the solar system around the galactic center, approximately in the direction of the Cygnus constellation. Directional detection proposes to use the anisotropy in the angular distribution of the recoils arising from the motion of the Earth around the galactic center to get an unambiguous signature of the WIMP signal~\cite{Spergel1988}.
	
	\vspace{2mm}
	In the next section, we introduce the MIMAC detector.
	As a directional detector, MIMAC provides observables to retrieve both the kinetic energy and the initial direction of the WIMP-induced recoiling ions:
	we detail how the kinetic energy is determined in section \ref{sec:energy} and present ongoing studies of the directionality in section \ref{sec:headtail}. 
	In the last section, we briefly discuss the ongoing developments and mention the comparison with other directional strategies.

	\vspace{-3mm}
	\section{The MIMAC detector}
	\label{sec:mimac}
	
	\vspace{-3mm}
	The MIMAC (MIcro-tpc MAtrix of Chambers) collaboration has developed an original directional detector~\cite{Santos2007}.
	The idea is to use a low-pressure gaseous TPC to improve the electron/nuclear recoils discrimination and to observe tracks long enough  to provide the sought for direction of the recoils.
	A specific gas mixture has been developed in order to precisely control the gain and the velocity of the primary electrons in the drift chamber : $70\%$ $\textrm{C}\textrm{F}_4$+$28\%$ $\textrm{C}\textrm{H}\textrm{F}_3$+$2\%$ $\textrm{C}_4\textrm{H}_{10}$.
	The few dozens of primary electrons produced by ionization along the recoil track are collected by an electric field of 180 V/cm toward the anode. These electrons  are amplified at the anode level by a 256\,$\mu$m-gap pixelated Micromegas~\cite{Iguaz} which has been specifically designed for a working pressure of 50\,mbar. The Micromegas has 256$\times$256 pixels placed on 424\,$\mu$m wide pitches in both X and Y directions, leading to a 10.8\,cm\,$\times$\,10.8\,cm total area.
	The signal is retrieved by a total of 512 channels -- 256 in X-axis and 256 in Y-axis; each of these has its own threshold above the background noise. A coincidence between a X and a Y strip/channel defines a triggered pixel. An additional channel measures the integrated signal on the grid of the Micromegas, over the whole detecting area, to retrieve the ionization energy of the events.
	A self-triggered electronic system~\cite{Richer} has been developed at LPSC, and allows a sampling of the channels at 50\,MHz. This provides a 3D description of the track -- the depth Z is defined by blocks of 20\,ns (``timeslices''). 
	
	In order to discriminate the background events at the cathode level
	and to optimize the chamber volume, two detectors are combined, facing each other in a single gas vessel and sharing a common, 12\,$\mu$m thin, aluminized mylar cathode, at 25\,cm from each Micromegas grid.
	A first two-chamber module was installed in the Modane Underground Laboratory (LSM) in June 2012, followed by upgrades in June 2013 and June 2014. The stability of the gas quality is insured by a circulation system composed of a pump, renewing the bichamber gas volume every hour, a buffer volume, a pressure regulator, an oxygen and a radon filter.
	
	The discrimination between electronic and nuclear recoils is obtained by applying selection cuts on the different MIMAC observables. For instance, the pulse shape from the preamplifier is more asymmetric and presents a longer rise-time for electron events; at the track level, it has been observed that electrons lead to more diluted ionization electron clouds, leading to ``holes'' in the track reconstruction. A selection of the symmetric charge pulses and denser track will increase the nuclear-over-electron recoils ratio. 
	A multivariate analysis has been implemented~\cite{Billard_discri,Riffard_discri}, making the best use of the over 20 available MIMAC observables: a boosted decision tree was trained on data taken using a monochromatic neutron field. The electron rejection power of this technique was estimated on simulations: from 10$^2$ with a 99.8\% nuclear recoil efficiency to 10$^5$ with a 85.1\% nuclear recoil efficiency, on the full energy range.
	
	Data taken in LSM have been analyzed in order to characterize the $^{222}$Rn progeny coming from intrinsic material pollution~\cite{Riffard_radon}. Cuts on the MIMAC observables were made to select the events measured in coincidence on both sub-chambers, involving only one $\alpha$ particle saturating the preamplifier, and with the largest transverse diffusion. This provided the ionization energy spectrum of $^{214}$Pb and $^{210}$Pb daughters emitted at the cathode level, along with  the track information of each such Pb nucleus.
	
	\vspace{-3.7mm}
	\section{Energy calibration}
	\label{sec:energy}
	
	\vspace{-3.7mm}
	A charge preamplifier connected to the grid of the Micromegas delivers the charge integration on the whole anode area every 20\,ns. The total ionization energy of an event is the baseline-subtracted integration over the total duration of the track, given in terms of ADC units. 
	Two steps are needed to retrieve the kinetic energy of the event.
	
	The \textit{ADC-to-keV parameters} are derived from measurements of electrons with a known kinetic energy. Electrons above 100\,eV  indeed leave all their kinetic energy in ionization~\cite{Waker}. 
	In Modane, this calibration is performed by producing, in the two chambers, fluorescence photons from cadmium (3.2\,keV), iron (6.4\,keV) and copper (8.1\,keV) foils bombarded by an X-ray generator. This calibration is done weekly and has shown a remarkable stability of the gas quality for more than two years of operation with the circulating pump.
	
	Unlike electrons, ions do not leave all their energy in the ionization channel. Hence the need of a specific study of the \textit{ionization quenching factor} (IQF), \ie~the proportion of the kinetic energy released in ionization.
	This proportion depends on the considered nucleus, on the value of its initial kinetic energy, on the composition and pressure of the gas mixture. 
	Analytical~\cite{Lindhard} and semi-analytical~\cite{Ziegler,Ziegler_2} models proposing quenching factors for a given configuration exist, but previous measurements have shown that these models fail to reconcile with data at energies below 50\,keV~\cite{santos,Guillaudin}.
	The only way to retrieve the IQF is to measure it as a function of the kinetic energy for each specific configuration. A portable ion beam facility has been designed at LPSC to make IQF measurements: COMIMAC~\cite{Muraz}. It uses a COMIC source~\cite{sortais} to produce both ions and electrons beams up to kinetic energies of 50\,keV. In ion configuration, a Wien filter makes for an efficient charge-over-mass separation.
	As such, it can both be used to calibrate the ADC-to-keV coefficients -- in electron mode -- and measure the IQF -- in ion mode. Although primarily designed to measure the IQF for MIMAC related gas mixture and ion targets, it can be used to measure the IQF in any gas mixture for any pressure up to 10\,bar.
	
	\vspace{-3.7mm}
	\section{Reconstruction of tracks with a controlled direction and energy}
	\label{sec:headtail}
	
\begin{figure}
		\centering
		\includegraphics[width=0.74\linewidth, trim={0 0mm 0 5mm},clip]{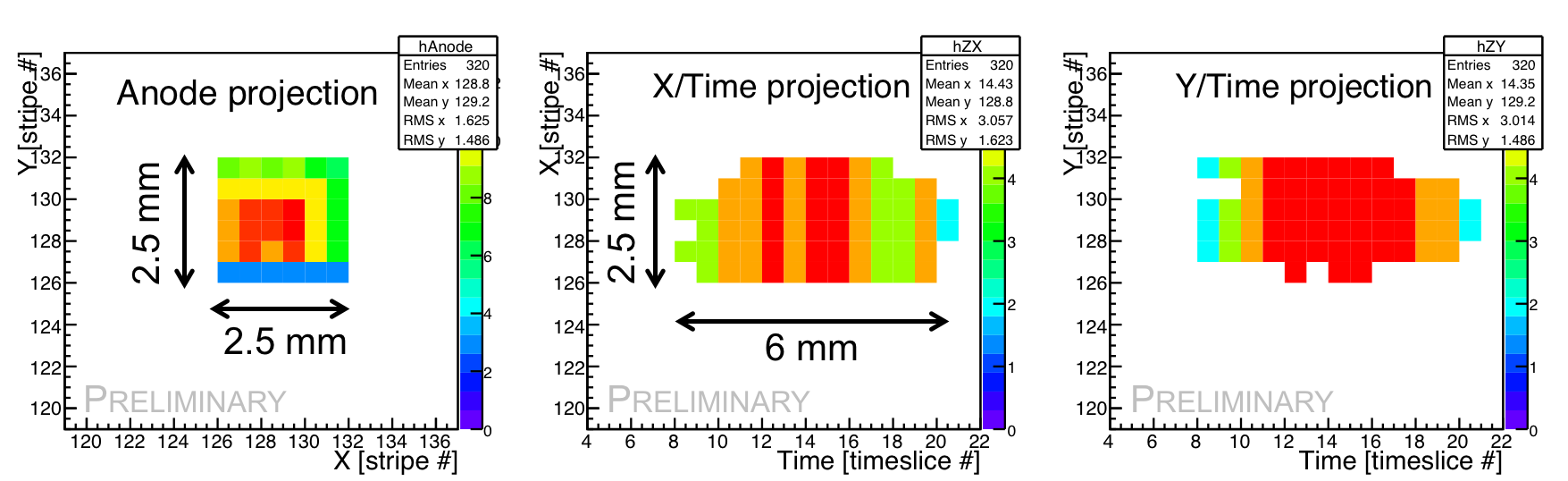}
		\includegraphics[width=0.24\linewidth, trim={0 0 0 0},clip]{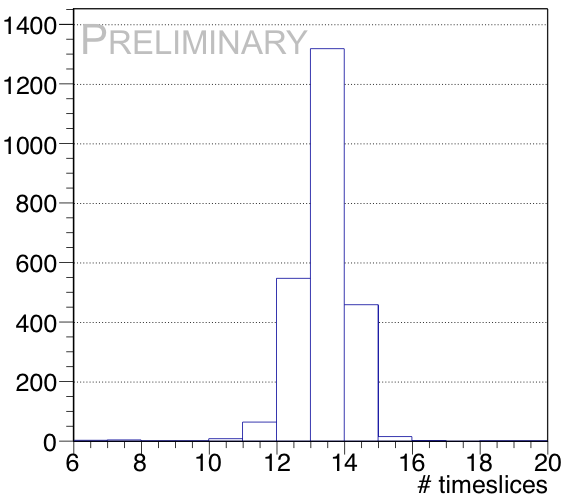}
		
		\footnotesize
		\hspace{4.2cm} (a) \hspace{7.5cm} (b)
		
		\vspace{-2mm}
		\caption{(a) Projection of the track of a $^{19}$F$^+$ ion sent by COMIMAC with a controled kinetic energy of 25\,keV into a MIMAC chamber filled with  $70\%$ $\textrm{C}\textrm{F}_4$+$28\%$ $\textrm{C}\textrm{H}\textrm{F}_3$+$2\%$ $\textrm{C}_4\textrm{H}_{10}$ at 50\,mbar. (b) Distribution of the number of timeslices (\ie~multiples of 20\,ns) of 25\,keV kinetic energy fluorine tracks. The maximum occurrence value of 13\,timeslices corresponds to 6.5\,mm. \vspace{-3mm}}
		\label{fig:track}
	\end{figure}

	\vspace{-3.7mm}
	In order to characterize the 3D tracks of low energy ions in the MIMAC gas mixture, the COMIMAC line was coupled to a MIMAC detector at LPSC. The two chambers are connected by a 1\,$\mu$m diameter hole which limits, at the molecular level, the leakage from the TPC (50\,mbar) into COMIMAC ($10^{-5}$ mbar). 
	Several data taking campaigns have been conducted to send electrons, protons and $^{19}\textrm{F}$  in the MIMAC gas mixture with energies from a few keV to 12\,keV (electrons) and 30\,keV (fluorine ions). We have been able to measure tracks of fluorine ions down to 5\,keV. Figure \ref{fig:track}a shows the projection of a track of a $^{19}$F$^+$ ion sent by COMIMAC with a controled kinetic energy of 25\,keV into a MIMAC chamber; we measured an ionization energy of 9\,keV for this event. The distribution of the number of timeslices of 25\,keV kinetic energy fluorine tracks is presented in Figure \ref{fig:track}b; the maximum occurrence corresponds to 6.5\,mm tracks.  Going to lower energies is possible, yet the lower part of the energy spectrum is filled with heavier species. Also, the gain is very sensitive to the amount of pollution of the gas. Further optimizations and data taking are ongoing to try to observe $^{19}\textrm{F}$ tracks down to 1\,keV. 
	The performance of the MIMAC reconstruction, \eg~in terms of track length and angular distribution as a function of the energy, will be detailed in a separate publication~\cite{couturier_comimac}.
	
	\vspace{-3mm}
	\section{Perspective: Ongoing R\&D and comparison with other strategies}
	
	\vspace{-3mm}
	The prototype presented in section \ref{sec:mimac} is an element of the final matrix. 
	Developments are going on to design high radio purity, low cost and larger (35$\times$35\,cm$^2$) Micromegas detectors. The next step of 1\,m$^3$ of effective volume, to be installed at LSM, will thus require only 16 bi-chamber modules, assembled in one big vessel. The data acquisition framework has had a major update, and it is now able to manage the multiple modules required for the \,m$^3$-scale extension.
	
	Other directional techniques have been proposed, \eg~using emulsion~\cite{Naka} or crystal~\cite{Sekiya,cappella} detectors. 
	Monte-Carlo simulations can be used to emulate the motion of an ion recoiling due to the elastic scattering by a DM particle in different sensing materials. A new  figure of merit is proposed to measure the preservation of the initial direction information in a given detecting material: the average over the entire track of the cosines of the angles between the initial direction of recoil and the direction after each collision. 
	This observable could help comparing the different directional strategies as regards the measurement of the direction of WIMP-induced nuclear recoils~\cite{couturier_comparison}.
	
	\vspace{2mm}
	CC acknowledges support from the Labex ENIGMASS.

\end{document}